\documentclass[Physsubmission, Phys]{SciPost}

\hypersetup{
    colorlinks,
    linkcolor={red!50!black},
    citecolor={blue!50!black},
    urlcolor={blue!80!black}
}

\usepackage[bitstream-charter]{mathdesign}
\newcommand{\lsim}{\raisebox{-0.13cm}{~\shortstack{$<$ \\[-0.07cm]
      $\sim$}}~}

\newcommand{\gsim}{\raisebox{-0.13cm}{~\shortstack{$>$ \\[-0.07cm]
      $\sim$}}~}

\DeclareSymbolFont{usualmathcal}{OMS}{cmsy}{m}{n}
\DeclareSymbolFontAlphabet{\mathcal}{usualmathcal}

\begin{document}


\begin{center}{\Large \textbf{
The Status of the Galactic Center Gamma-Ray Excess \\
}}\end{center}


\begin{center}
Dan Hooper 
\end{center}

\begin{center}
Fermi National Accelerator Laboratory\\
University of Chicago
\\
* dhooper@fnal.gov
\end{center}

\begin{center}
\today
\end{center}


\definecolor{palegray}{gray}{0.95}
\begin{center}
\colorbox{palegray}{
  \begin{tabular}{rr}
  \begin{minipage}{0.1\textwidth}
    \includegraphics[width=30mm]{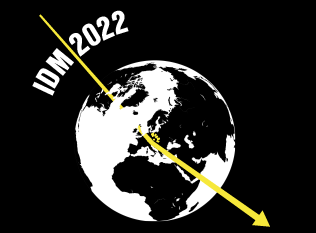}
  \end{minipage}
  &
  \begin{minipage}{0.85\textwidth}
    \begin{center}
    {\it 14th International Conference on Identification of Dark Matter}\\
    {\it Vienna, Austria, 18-22 July 2022} \\
    \doi{10.21468/SciPostPhysProc.?}\\
    \end{center}
  \end{minipage}
\end{tabular}
}
\end{center}

\section*{Abstract}
{\bf
The Galactic Center Gamma-Ray Excess has a spectrum, angular distribution, and overall intensity that agree remarkably well with that expected from annihilating dark matter particles in the form of a $m_X \sim 50 \, {\rm GeV}$ thermal relic. Previous claims that these photons are clustered on small angular scales or trace the distribution of known stellar populations once appeared to favor interpretations in which this signal originates from a large population of unresolved millisecond pulsars. More recent work, however, has overturned these conclusions, finding that the observed gamma-ray excess does {\it not} contain discernible small scale power, and is distributed with approximate spherical symmetry, not tracing any known stellar populations. In light of these results, it now appears significantly more likely that the Galactic Center Gamma-Ray Excess is produced by annihilating dark matter.
%
}

\section{Introduction}
\label{sec:intro}

An excess of GeV-scale gamma-rays from the region surrounding the Galactic Center was discovered in the publicly available data collected by the Fermi Telescope over 12 years ago~\cite{Goodenough:2009gk,Hooper:2010mq,Hooper:2011ti}. In the years that followed, the detailed characteristics of this signal were measured with increasing precision~\cite{Daylan:2014rsa,Calore:2014xka,Calore:2014nla,Cholis:2021rpp}, although the basic spectral and morphological features of the excess remained unchanged. More specifically, these studies found that the spectrum, angular distribution, and overall intensity of this signal are each in good agreement with the predictions of annihilating dark matter in the form of a $m_X \sim 50 \, {\rm GeV}$ thermal relic.

During a brief period in 2014-15, there was a particularly high degree of excitement around the possibility that Fermi may have detected dark matter annihilation products. This enthusiasm fell precipitously in 2015, however, when two independent groups, using two different analysis techniques, reported that they had found evidence that the photons constituting this excess are spatially clustered, suggesting that they originate from a population of near-threshold astrophysical point-sources (such as millisecond pulsars), rather than from annihilating dark matter~\cite{Lee:2015fea,Bartels:2015aea}. Further advancing this conclusion, multiple groups, starting in 2016, reported that the angular distribution of this excess is not spherical (as would be expected from dark matter), but instead is correlated with the distribution of stars that make up the Galactic Bulge and Bar~\cite{Bartels:2017vsx,Macias:2016nev,Macias:2019omb,Pohl:2022nnd}.

In this proceeding, I will provide an important update of this situation. In particular, I will describe how recent work has determined that the gamma-ray excess {\it is} approximately smooth and spherically symmetric, with claims to the contrary having been shown to be spurious. The best present assessment of the Fermi data is that it contains an excess of GeV-scale emission that is spatially smooth and spherically symmetric, to the best of our ability to currently measure. In light of this, dark matter interpretations of this signal appear much more likely than they did as recently as a few years ago. The challenges involved in trying to explain this signal with millisecond pulsars, in contrast, have become only greater as new information has come to light.

\section{The Excess is Spatially Smooth}

In standard analyses of Fermi data, the angular distribution of the photons in a given energy bin is compared to the sum of a collection of spatial templates, whose coefficients are varied until the best fit (and the uncertainty around the best fit) is identified. In most situations, the likelihood of observing a given number of events in a spatial bin is taken to be equal to the Poisson probability, as calculated using the mean value associated with the templates in question. This procedure is appropriate for most templates, such as those associated with pion production, bremsstrahlung, inverse Compton scattering, the Fermi Bubbles, the extragalactic gamma-ray background, and annihilating dark matter. If the Galactic Center Gamma-Ray Excess is generated by a population of near-threshold point sources, however, the number of photons in a given spatial bin might not follow a Poisson distribution, but instead would follow a distribution that is related to the luminosity function of the point source population in question. With this possibility in mind, the authors of Ref.~\cite{Lee:2015fea} conducted an analysis of the Fermi data employing a combination of Poissonian and non-Poissonian templates, finding a strong statistical preference for the excess to be attributed to a non-Poissonian template, suggesting that this signal arises from a population of point sources (such as millisecond pulsars) which are only slightly below Fermi's current detection threshold.

While the authors of Ref.~\cite{Lee:2015fea} performed many tests of their analysis pipeline at the time, and otherwise took great efforts to validate their results, it was ultimately shown that their conclusions were an artifact of insufficiently understood backgrounds. This was clearly demonstrated by Leane and Slatyer in 2019~\cite{Leane:2019xiy} (see also, Refs.~\cite{Leane:2020pfc,Leane:2020nmi}) who repeated the non-Poissonian analysis of Ref.~\cite{Lee:2015fea}, but then tested the validity of that study's conclusions by performing an injection test. More specifically, they found that when they added a simulated dark matter-like (i.e., smooth) signal to the real Fermi data, the fit strongly preferred to attribute the simulated photons to a non-Poissonian template. In other words, even when the signal in question was a smooth by design, the non-Poissonian template fit strongly preferred to incorrectly identify it as clumpy. 

Another study, using a technique involving spatial wavelets, also claimed to show that the Galactic Center Gamma-Ray Excess arises from a population of near-threshold point sources~\cite{Bartels:2015aea}. This conclusion, however, was shown to be incorrect in 2019 by Zhong, McDermott, Cholis and Fox~\cite{Zhong:2019ycb}, who repeated the wavelet-based analysis using an updated catalog of gamma-ray point sources. When utilizing the updated catalog, no significant evidence of small-scale power was found among the photons that make up the Galactic Center Gamma-Ray Excess.

To be clear, the non-Poissonian and wavelet-based analysis techniques are each mathematically sound under appropriate conditions, and the analyses of Refs.~\cite{Lee:2015fea} and~\cite{Bartels:2015aea} were correctly identifying small-scale power in the observed distribution of photons. The subsequent analyses of Refs.~\cite{Leane:2019xiy} and~~\cite{Zhong:2019ycb}, however, showed that this clustering was taking place among the astrophysical backgrounds, and could not be attributed to the Galactic Center Gamma-Ray Excess. At present, this class of analyses can only be used to place an upper limit on the fraction of the excess that originates from sources with a flux above a given value, thus allowing one to place constraints on the luminosity function of any point source population that might be responsible for this signal (see Sec.~\ref{sec:strain}).

\begin{figure}[t]
\centering
\includegraphics[width=0.32\textwidth]{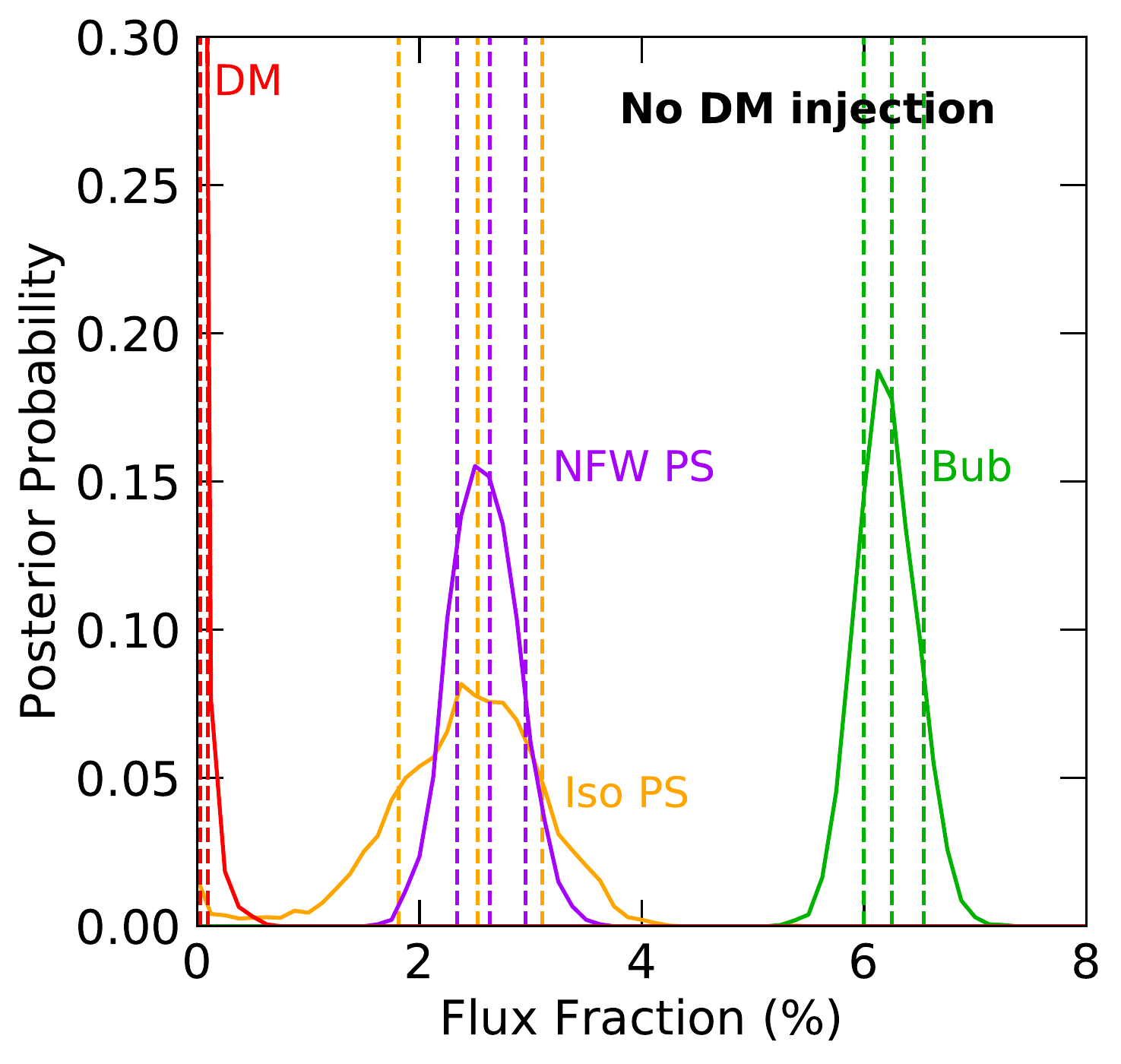}
\includegraphics[width=0.32\textwidth]{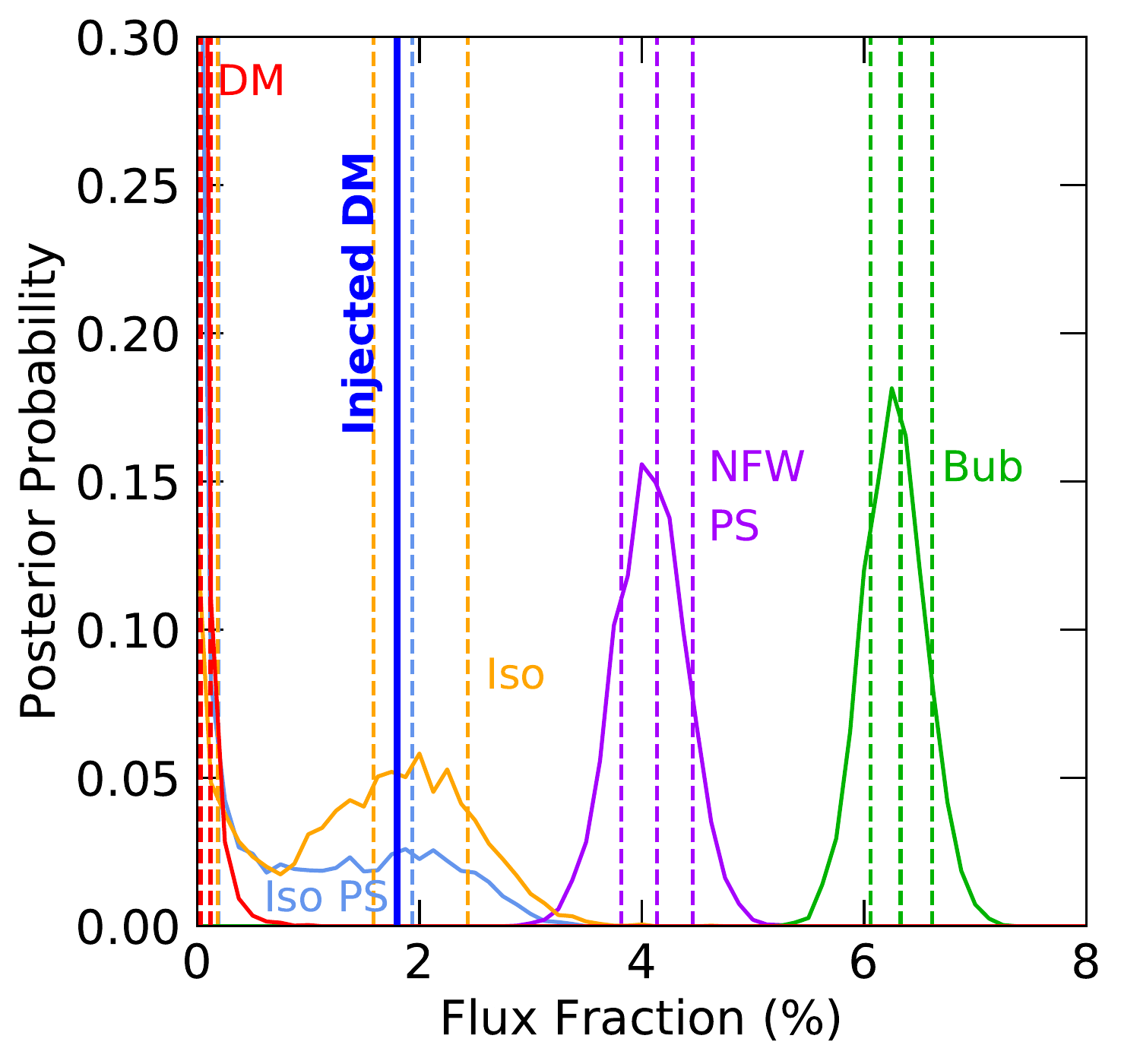}
\includegraphics[width=0.32\textwidth]{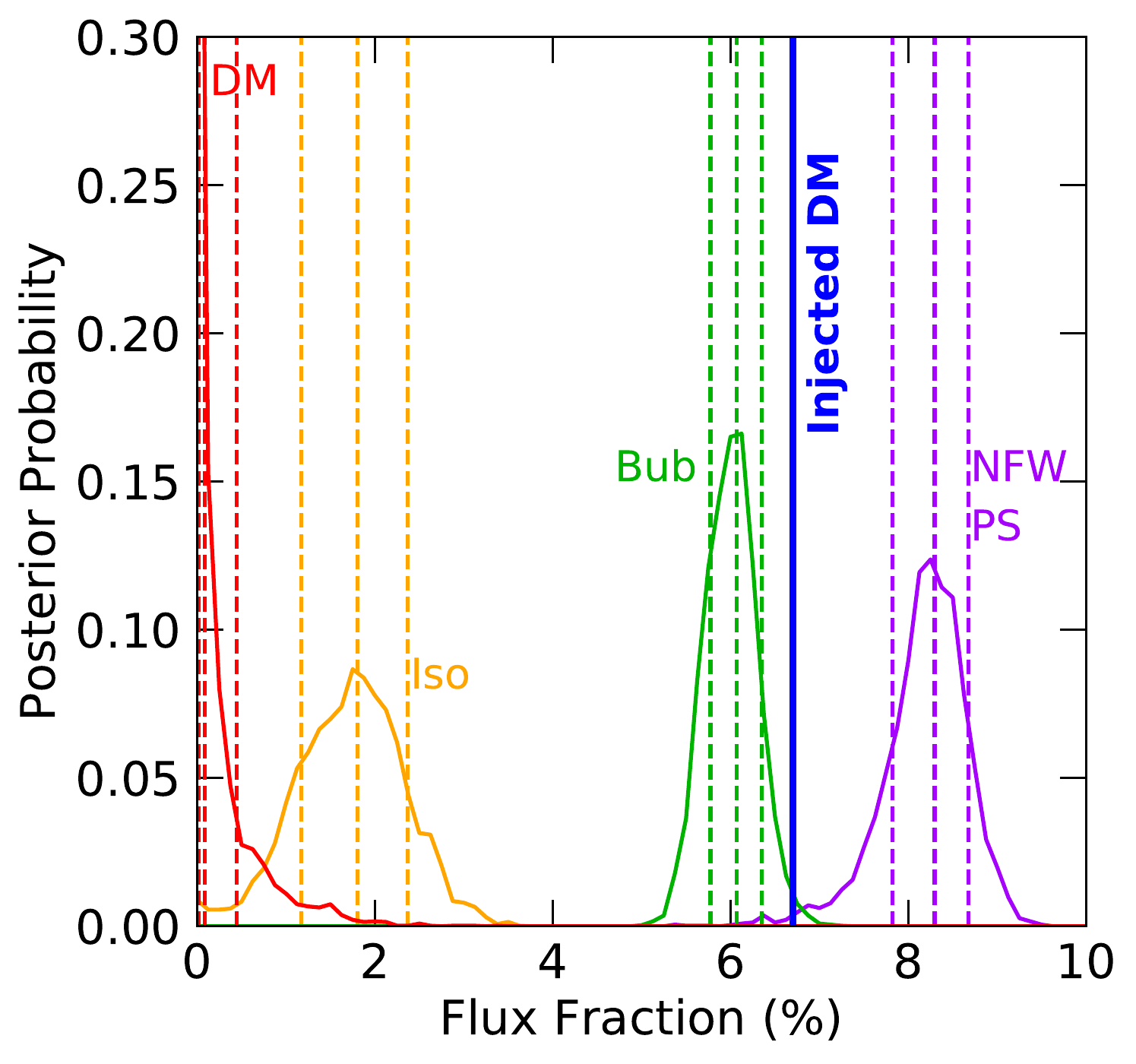}
\caption{The fraction of the Galactic Center Gamma-Ray Excess that is attributed to annihilating dark matter (``DM'') or to unresolved point sources (``NFW PS'') in a non-Poisson template fit. In the left frame, the real Fermi data has been used, while in the center and right frames a simulated dark matter-like (i.e., smooth) signal has been added to the real Fermi data. Despite having introduced a smooth component of excess emission, the non-Poissonian template fit strongly prefers to incorrectly identify this signal as clumpy. From Ref.~\cite{Leane:2019xiy}.}
\label{ref}
\end{figure}

\section{The Excess is Spherically Symmetric}

Since 2016, a number of papers have claimed that better fits to the Fermi data are obtained when the excess is modeled using a combination of templates which trace various known stellar populations, such as the Milky Way's Box-Shaped (or ``Boxy'') Bulge, X-Shaped Bulge, and Nuclear Bulge~\cite{Bartels:2017vsx,Macias:2016nev,Macias:2019omb,Pohl:2022nnd}. If true, this result would favor astrophysical interpretations of the gamma-ray excess. In the first of these studies, the authors reported that the excess traces a template associated with the X-Shaped Bulge~\cite{Macias:2016nev}. This result, however, was later shown to be an artifact of the relatively small region-of-interest that was adopted in that study~\cite{Bartels:2017vsx}. In the more recent works of Refs.~\cite{Macias:2019omb,Pohl:2022nnd}, the authors claimed that the excess is better fit by a combination of spatial templates associated with the Boxy Bulge and Nuclear Bulge. This is in stark disagreement with the conclusions reached by Di Mauro~\cite{DiMauro:2021raz} and by Cholis, Zhong, McDermott and Surdutovich~\cite{Cholis:2021rpp}, who find that the excess is much better fit by a spherical (i.e., dark matter-like) template.

Until recently, it was not obvious (at least to me) which of these very different conclusions was more likely to be correct. It has since become significantly more clear, however, that the excess is, in fact, consistent with having a spherical (i.e., dark matter-like) morphology and does not significantly correlate with any known stellar populations. In particular, the fits based on the templates from Refs.~\cite{Macias:2019omb,Pohl:2022nnd} are far worse (at a level of $\Delta \chi^2 \sim 6500$) than those based on the templates from Cholis et al.~\cite{Cholis:2021rpp}, despite involving a larger number of degrees-of-freedom~\cite{McDermott:2022zmq}. In other words, the fits only prefer the excess to have a bulge-like morphology when the background model provides a poor fit to the overall dataset. Whenever the model yields a high value for the overall likelihood, the excess is always strongly preferred to be spherical. It also appears possible that the analyses pipelines used in Refs.~\cite{Macias:2019omb,Pohl:2022nnd} may have failed to identify the global minimum of the likelihood, and are instead comparing fits with spherical and bulge-like excesses at a {\it local} minima, leading to spurious results~\cite{McDermott:2022zmq}.

\begin{figure}[t]
\centering
\includegraphics[width=0.6\textwidth]{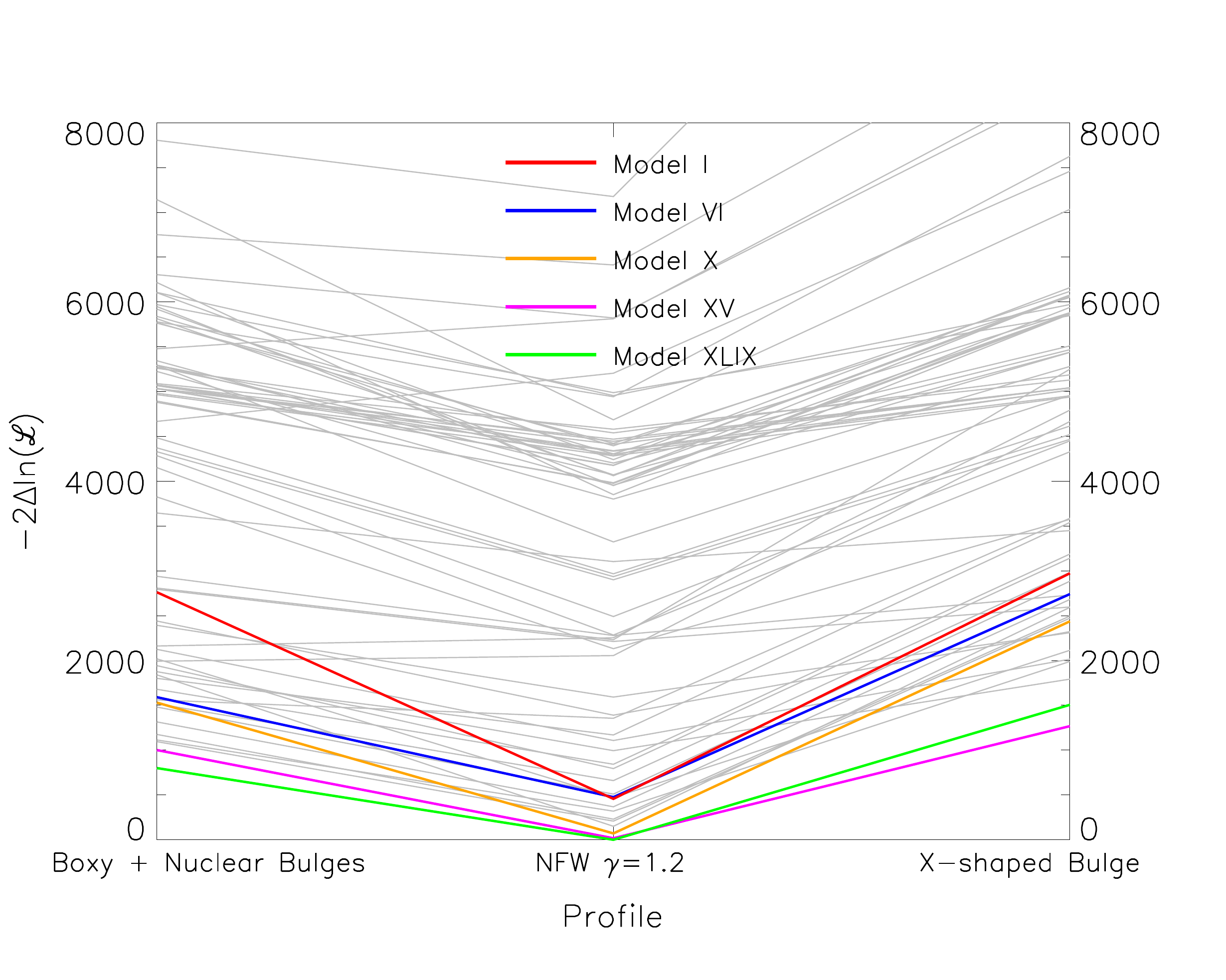}
\caption{A comparison of the log-likelihoods obtained when the gamma-ray excess is modeled as being spherically symmetric and dark matter-like (center), or as tracing a sum of the Boxy and Nuclear Bulges (left), or the X-Shaped Bulge (right). For all models that provide a reasonably good fit to the Fermi data, a dark matter-like morphology is strongly preferred over those tracing known stellar populations. From Ref.~\cite{Cholis:2021rpp}.}
\end{figure}

\section{Pulsar Interpretations are Strained on Several Fronts}
\label{sec:strain}

As described in the previous two sections, the Fermi data show no signs that the Galactic Center Gamma-Ray Excess is clumpy, or that it traces any known stellar populations. That being said, these results do not preclude the possibility that this signal could be produced by a large population of very faint gamma-ray point sources that are distributed with approximate spherical symmetry around the Galactic Center. Since pulsars are the only known class of astrophysical objects that produce a gamma-ray spectrum with a similar shape to that the Galactic Center Gamma-Ray Excess, they are often the focus of such discussions. More specifically, the long lifetimes of millisecond (or ``recycled'') pulsars make them the primary astrophysical candidate for the origin of the gamma-ray excess.

There are several reasons, however, to conclude that millisecond pulsars are unlikely to produce much of the observed excess. First of all, the gamma-ray luminosity function of the millisecond pulsar populations found in globular clusters and in the Galactic Disk have been measured to peak near $L_{\gamma} \sim 10^{34}-10^{35} \, {\rm erg/s}$ (in $L^2 dN/dL$ units)~\cite{Hooper:2016rap,Gonthier:2018ymi,Bartels:2018xom,Cholis:2014noa,Hooper:2015jlu}. If the gamma-ray excess is generated by a population of pulsars with a similar luminosity function, Fermi should have already detected $\sim 10^2$ such sources at high significance~\cite{Dinsmore:2021nip}. As of this time, however, no millisecond pulsars have been detected near the Galactic Center.\footnote{In Ref.~\cite{Ploeg:2020jeh}, the authors claimed that the millisecond pulsars PSR J1747-4036, PSR J1811-2405 and PSR J1855-1436 are likely to be members of an Inner Galaxy population, but the measured distances to these pulsars rules out this possibility~\cite{2017ApJ...835...29Y}.} Furthermore, it was shown in Ref.~\cite{Bartels:2017xba} that masking all of the gamma-ray sources (and source candidates) which exhibit a pulsar-like spectrum does not impact the measured spectrum or intensity of the excess. From this, we can conclude that if millisecond pulsars (or other point sources) generate the excess, they must be very numerous ($\gsim 10^4-10^5$) and significantly less luminous than millisecond pulsars found elsewhere, featuring a luminosity function that peaks at $L_{\gamma} \lsim 10^{33} \, {\rm erg/s}$.

Millisecond pulsars are formed when they are ``spun up'' by a binary companion, and the evolutionary precursors to millisecond pulsars are objects known as low-mass X-ray binaries. As pointed out in Ref.~\cite{Cholis:2014noa}, one can combine measurements of the gamma-ray emission from globular clusters, the number of bright low-mass X-ray binaries in globular clusters, and the number of bright low-mass X-ray binaries in the Inner Galaxy, to estimate the fraction of the gamma-ray excess that originates from millisecond pulsars. When this exercise was carried out in Ref.~\cite{Haggard:2017lyq}, it was shown that if the entire excess originates from millisecond pulsars, INTEGRAL should have observed $\sim 10^3$ bright low-mass X-ray binaries in the Inner Galaxy, whereas it actually detected only 42. This allows us to conclude that only $\sim 4-11\%$ of the excess can be attributed to millisecond pulsars. 

Finally, observations using the HAWC and LHAASO telescopes have shown that young and middle-aged pulsars are approximately universally surrounded by bright, spatially-extended, multi-TeV emitting regions known as ``TeV halos''~\cite{Hooper:2017gtd,Linden:2017vvb,HAWC:2021dtl,HAWC:2019tcx}. This emission is produced through the inverse Compton scattering of very high-energy electrons and positrons, and the intensity of the observed emission requires that $\mathcal{O}(10\%)$ of the pulsars' total energy budget (i.e., spindown power) goes into the acceleration of such particles~\cite{Hooper:2017gtd,Sudoh:2021avj}. It has recently been shown (at the 99\% C.L.) that millisecond pulsars also generate TeV halos, with an efficiency that is similar to that of young and middle-aged pulsars~\cite{Hooper:2021kyp}. If this result is robustly confirmed, we could use ground-based gamma-ray telescopes to search for the multi-TeV emission from TeV halos near the Galactic Center, providing us with an independent measurement of (or at least an independent upper limit on) our Inner Galaxy's millisecond pulsar population~\cite{Hooper:2018fih}. At present, measurements of the Milky Way's innermost $0.5^{\circ}$ by HESS~\cite{HESS:2016pst} are in modest tension with pulsar interpretations of the gamma-ray excess~\cite{Hooper:2018fih}. Future measurements of the inner several degrees around the Galactic Center by CTA will provide a powerful probe of the Inner Galaxy's pulsar population, and could supply a definitive test of whether pulsars are responsible for the observed gamma-ray excess~\cite{inprep}.

\section{Conclusion}

The gamma-ray emission observed from the region surrounding the Galactic Center includes an excess relative to known astrophysical sources and emission mechanisms. Furthermore, the spectrum, angular distribution, and overall intensity of this excess are consistent with arising from the annihilations of a $m_X \sim 50 \, {\rm GeV}$ dark matter particle with an annihilation cross section of $\langle \sigma v \rangle \sim 10^{-26} \, {\rm cm}^3/{\rm s}$, in good agreement with the expectations of a thermal relic. In contrast to previous claims, it has recently become clear that this signal does {\it not} show evidence of small scale power, and does {\it not} trace any known stellar populations. Instead, the gamma-ray excess is approximately smooth and distributed with spherical symmetry around the Galactic Center, as would be expected of dark matter annihilation products.

While it remains possible that this signal could be generated by a large population of very faint gamma-ray point sources (such as millisecond pulsars), distributed with approximate spherical symmetry around the Galactic Center, this seems unlikely for several reasons. In particular, in order to explain the fact that Fermi has not detected many bright pulsar candidates from the direction of the Inner Galaxy, the luminosity function of any source population that could potentially be responsible for the excess must peak at a significantly lower value than is measured among the millisecond pulsar populations observed in globular clusters or the Galactic Disk. Furthermore, if millisecond pulsars were responsible for the gamma-ray excess, we should have observed many more bright low-mass X-ray binaries in the Inner Galaxy, as well as a greater intensity of TeV-scale emission. From these and other considerations, we can conclude that if pulsars are, in fact, responsible for this signal, the Inner Galaxy's pulsar population must have very different properties than those observed elsewhere in the Milky Way. 

Looking to the future, radio searches for millisecond pulsars in the Inner Galaxy~\cite{Calore:2015bsx}, as well as measurements of the Inner Galaxy with CTA~\cite{inprep}, will each provide important constraints on the Inner Galaxy's pulsar population. In addition, searches for dark matter annihilation products from dwarf galaxies, and in the anti-proton and anti-nuclei components of the cosmic-ray spectrum, will play an important role in confirming or constraining dark matter interpretations of the Galactic Center Gamma-Ray Excess.

\section*{Acknowledgements}

I would like to thank Yiming Zhong, Anastasia Sokolenko, and Celeste Keith for helpful comments. I would also like to thank the local organizers of IDM 2022 for organizing an outstanding edition of this conference series. This work has been supported by the Fermi Research Alliance, LLC under Contract No.~DE-AC02-07CH11359 with the U.S. Department of Energy, Office of High Energy Physics.



\bibliography{IDM_Hooper.bib}

\nolinenumbers

\end{document}